\documentclass[]{article}

\usepackage{graphicx}
\usepackage{amsmath,amssymb}
\usepackage{bm}
\usepackage{rotating}
\usepackage{authblk}
\usepackage[shortcuts]{extdash}

\title{Evaluation of Performance for Human In-Vivo Conductivity Estimation from EEG and sEEG Recorded in Simultaneous with Intracerebral Electrical Stimulation}
\author[1,2]{Hamza Altakroury}
\author[1,2,3]{Laurent Koessler}
\author[1,2]{Radu Ranta}
\author[4]{Janis Hofmanis}
\author[1,2,3]{Sophie Colnat Coulbois}
\author[1,2,3]{Louis Maillard}
\author[1,2]{Valérie Louis Dorr}
\affil[1]{Université de Lorraine, CRAN UMR 7039, Campus Science BP 70239, Vandoeuvre-les-Nancy Cedex, 54506, France}
\affil[2]{CNRS, CRAN UMR 7039, France}
\affil[3]{Service de Neurologie, Unit d’épileptologie, Centre Hospitalier Regional et Universitaire de Nancy, Nancy, 54000, France}
\affil[4]{Ventsplis University, Ventsplis Engineering Research Institute, Ventsplis, LV3601, Latvia}
\date{}                     
\begin{document}

\bibliographystyle{unsrt}

\maketitle

\noindent\rule{\textwidth}{1pt}

\section*{Abstract}
Assigning accurate conductivity values in human head models is an essential factor for performing precise electroencephalographic (EEG) source localization and targeting of transcranial electrical stimulation (TES). Unfortunately, the literature reports diverging conductivity values of the different tissues in the human head. The current study analyzes first the performance of in-vivo conductivity estimation for different configurations concerning the localization of the electrical source and measurement. Then, it presents conductivity estimates for three epileptic patients using scalp EEG and intracerebral stereotactic EEG (sEEG) acquired in simultaneous with intracerebral electrical stimulation. The estimates of the conductivities were based on finite-element models of the human head with five tissue compartments of homogeneous and isotropic conductivities. The results of this study show that in-vivo conductivity estimation can lead to different estimated conductivities for the same patient when considering different stimulation positions, different measurement positions or different measurement modalities (sEEG or EEG). This work provides important guidelines to in-vivo conductivity estimation and explains the variability among the conductivity values which have been reported in previous studies. \\

Key words: Key words: EEG, Epilepsy, Finite element model, Head conductivities, Intracerebral electrical stimulation, sEEG.

\section{Introduction}
Nowadays, the reference method for localizing the epileptogenic zones is the stereo EEG (sEEG) measurements which are recorded by electrodes implanted inside the brain \cite{koessler2010source}. The sEEG presents two different functions: 1) Recording the spontaneous seizures and 2) Stimulating by intracerebral electrical stimulation (IES) for inducing seizures and determining the surrounding functional areas. However, the recent developments of reliable and useful non-invasive tools supporting the EEG Source Imaging (ESI) have urged the researchers to localize the epileptogenic zones non-invasively \cite{koessler2010source,plummer2008eeg,rikir2014electrical}, by solving the so-called inverse problem (estimating sources from distant measurements). Yet, for solving the inverse problem, an accurate forward head model describing the electrical propagation in a given head geometry with fixed conductivities needs to be identified.

The forward head model is a parametrized biophysical model linking the sources with the measured potentials. The inverse problem solution critically depends on the accuracy of the forward model, both in terms of source characteristics and in terms of propagation models. Determining the parameter values of the forward model is very complicated since the head compartments are anatomically complex and, in principle, inhomogeneous and anisotropic. Different modeling techniques exist in neuroscience, with different temporal and spatial resolutions \cite{frackowiak2015future}. Each model has its own view of the brain, for example, in connectivity studies the propagation occurs through the axons; while in conductivity estimation studies the propagation of the electrical and magnetic fields take place in every direction from the sources to the sensors. In addition, brain sources are complex to model due to their temporal dynamics and spatial organization. Yet, classical macroscopic source localization approaches are based on two main simplified hypothesis: 1) The sources can be modeled as dipole currents 2) The propagation can be tackled trough the quasi-static approximation of Maxwell equations (which holds for frequencies below 1 kHz \cite{nunez2006electric}). In this case, the forward problem of modeling the human head depends only on two important parameters: the geometry of the head model, and the conductivity of each compartment in the head model.

In earlier studies, the geometry of the head model was defined as one or more homogeneous spheres in which the output potentials can be determined analytically \cite{cuffin1991eccentric}. However, due to the improvement in imaging techniques and the computational capacity, realistic head models have been developed. Usually, anatomically realistic head models are generated from MR images of the head and the forward problem is solved numerically by different methods like the Boundary Element Method (BEM), the Finite Element Method (FEM) and the Finite Difference Method (FDM) \cite{hallez2007review}. Even though the FEM head model is becoming more common in recent studies, there is no proof of its superiority over the other numerical methods \cite{lew2009improved, adde2003symmetric, birot2014head} \footnote{Although the FEM models are highly flexible and allow anisotropic and inhomogeneous conductivities, they requires much more parameter tuning}. Several studies have shown that assigning accurate conductivity values for the different compartments in the head model has a significant effect on source localization, weather the geometry and the numerical method applied for generating the head model is simple or complex \cite{akalin2013effects, pohlmeier1997influence, homma1995conductivity}.

In the literature, brain conductivity estimation for humans or animals was performed generally by ex-vivo experiments \cite{geddes1967specific, rush1968current, baumann1997electrical}. Even though some of ex-vivo studies are old, their values still appear in recent articles \cite{ramon2004role, acar2010neuroelectromagnetic}. However, because the properties of tissues change when they are removed from their environment \cite{crile1922electrical}, in-vivo conductivity estimation has evolved. In-vivo conductivity estimation have been performed with different techniques like Electric Impedance Tomography (EIT) in which the brain is stimulated by scalp electrodes \cite{dabek2016determination, ferree2000regional, fernandez2011estimation, gonccalves2003vivo}. Using EIT, Gonçalves et al., performed a conductivity estimation of the scalp, the skull and the brain for six subjects. In their study, in which the scalp conductivity was assumed to be equal to the brain conductivity, the average of the resulted brain and skull conductivities over eight subjects were 0.33 ± 0.13 S/m and 0.0082 ± 0.18 S/m respectively \cite{gonccalves2003vivo}. Even though the EIT has a low cost and can be performed on healthy subjects, when applying the stimulation on the surface of the scalp, most of the current energy will be absorbed in the high-resistive skull before reaching the deep compartments of the brain (GM and WM).

Other works have performed conductivity estimation by considering the real sources of the brain \cite{lew2009improved, acar2016simultaneous}. In one study, Acar et al. performed an estimation of the brain-to-skull conductivity ratio in two subjects by considering the evoked responses from an arrow flanker task. In their study, the resulted brain-to-skull conductivity ratio was 34 for the first subject and 54 for the second \cite{acar2016simultaneous}. In a similar study, Lew et al. estimated the brain and the skull conductivity by considering the somatosensory evoked response for one subject as 0.48 S/m and 0.004 S/m \cite{lew2009improved} and therefore a ratio of 120. Although the conductivity estimation by brain sources is more realistic than EIT, since the current does not cross the high-resistive skull twice, these brain sources are not well-determined and their model may not be as accurate as modeling a well-determined source like the EIT. 

In other studies, in-vivo conductivity estimation has been performed by considering the subdural electrical stimulation \cite{zhang2006estimation, lai2005estimation}. Zhang et al. have estimated the brain-to-skull conductivity for two epileptic patients considering only the scalp EEG measurements and reported the average brain-to-skull conductivity ratio as 17.9 ± 2.3 and 19.9 ± 1 \cite{zhang2006estimation}. Lai et al. have considered both the scalp and the cortical potentials which were acquired in simultaneous with subdural stimulation and reported values between 18 and 34 for the brain-to-skull conductivity ratio for 5 epileptic patients \cite{lai2005estimation}. Although the subdural electrodes are more invasive than EIT for representing brain sources and better determined than physiological sources, they do not penetrate the cortex, so the effect of stimulating at different depths cannot be studied. In-vivo conductivity estimation with IES was performed by Koessler et al. \cite{koessler2017vivo}. However, in that work they considered a current with a much higher frequency (50 kHz current generated by RF generator) than the frequency of the EEG and the estimation was focused only on the gray matter and the white matter. The large variability in the conductivities estimated in previous studies is usually attributed to different estimation methods or differences in age, gender and health between subjects. Moreover, it is interesting to note that even though there are arguments that the tissues of the head are purely resistive \cite{huang2017measurements, ranta2017assessing}, some studies report that the distribution and attenuation of current may depend on the frequency when the estimation of brain conductivities is based on exogenous electrical sources \cite{opitz2016spatiotemporal}.

In this study, we perform in-vivo conductivity estimation in five-compartment FEM head models of epileptic patients with the sEEG and EEG recordings acquired during IES. The characteristics (position and amplitude) of exogenous intracerebral electrical sources (IES) are well known and this information allows to evaluate conductivity values using different measurement setups (sEEG or EEG) and different stimulation positions for the same patient. It is also important to note that IES is a periodic brief bipolar rectangular current pulse (see description in the next section and more detailed time and frequency analysis in \cite{ranta2017assessing, hofmanis2013denoising}). Beyond conductivity estimation, our aim is also to evaluate the sensitivity of these estimates depending on the source position relatively to the tissues of interest (scalp, skull, CSF, gray matter and white matter). The main question is whether the conductivity values of anatomically-detailed FEM models can be identified with more or less precision depending on various recording and stimulation configuration. The data used here is unique in that it combines intracranial voltage recording, intracranial electric stimulation and surface voltage recordings (in a brain considered normal from the point of view of electrical propagation). The objective of this work is thus to determine the impacts of both the position of the source of deep electrical stimulation and the spatial conditioning of the measurement electrodes on the estimation of the conductivities of the different anatomical structures of the head. In other words, given the energetic limitation of the stimulation currents, model errors, and the signal-to-noise ratio of the measurements, we aim to give a level of precision and therefore confidence in the estimates of the conductivities. In the following, the method for in-vivo conductivity estimation is first described in Section 2, then the results of simulation and real analyses are shown in Section 3. In Section 4, we discuss the results of this study and in section 5 we conclude the study.

\section{Materials and Methods}

\subsection{Subjects}

Three drug-resistant epileptic patient (Patient(1): male 23 years old, Patient(2): female 34 years old and Patient(3) female 21 years old) with partial epilepsy underwent simultaneous sEEG and EEG recordings combined with clinical routine IES. These patients were investigated in the Neurology Department of the University Hospital (CHU), Nancy. Each patient gave his informed consent and the study was approved by the ethical committee of the CHU Nancy.

Patients first underwent a presurgical evaluation including careful medical history examination, neurological examination, video-EEG recording (including seizures) and anatomic magnetic resonance imaging (MRI). Because several hypotheses were proposed by the noninvasive investigations, all patients underwent sEEG in order to delineate the epileptogenic zone. CT-scans were acquired after implanting the invasive electrodes in order to determine their positions in the brain of the patient \cite{salado2018seeg}. According to the noninvasive and invasive investigations the epileptogenic zones were in the left basal temporal and parahippocampal regions for Patient(1), in the right mesial temporal lobe for Patient(2) and in the right anterior insular cortex for Patient(3).

The average number of the intracerebral multi-contact electrodes was twelve (10 for Patient(1), 12 for Patient(2) and 14 for Patient(3)), with each electrode having eight to eighteen contacts. Equally spaced along the multi-contact electrodes, each sEEG contact is a cylinder of length 1.5 mm and diameter of 0.8 mm while the distance separating the centers of two adjacent contacts is 3.5 mm. The average number of scalp EEG electrodes was twenty-one (19 for Patient(1), 21 for Patient(2) and 24 for Patient(3)).

For defining the functionality of the surrounding areas of the epileptogenic zone, IES were performed according to clinical routine parameters \cite{trebuchon2016electrical}. Two adjacent contacts of a multi-contact invasive electrode were used to perform IES. The IES consists of periodic (55 Hz) biphasic rectangular current pulses with intensities between 0.2 mA and 2 mA, the width of the biphasic impulsion is 1.05 ms (brief, thus broad band). A complete clinical IES sequence lasts 5 seconds, but we only selected 2 seconds for our study in order to avoid commutation artifacts (\cite{hofmanis2013denoising}). The positions of the intracerebral electrodes were different for the different patients depending on the location of the epileptic zone. For the same patient, the stimulation was performed at different positions inside the brain, in order to perform a functional mapping.

\subsection{The forward model}

\begin{figure}[thpb]
\centering
\includegraphics[scale=0.4]{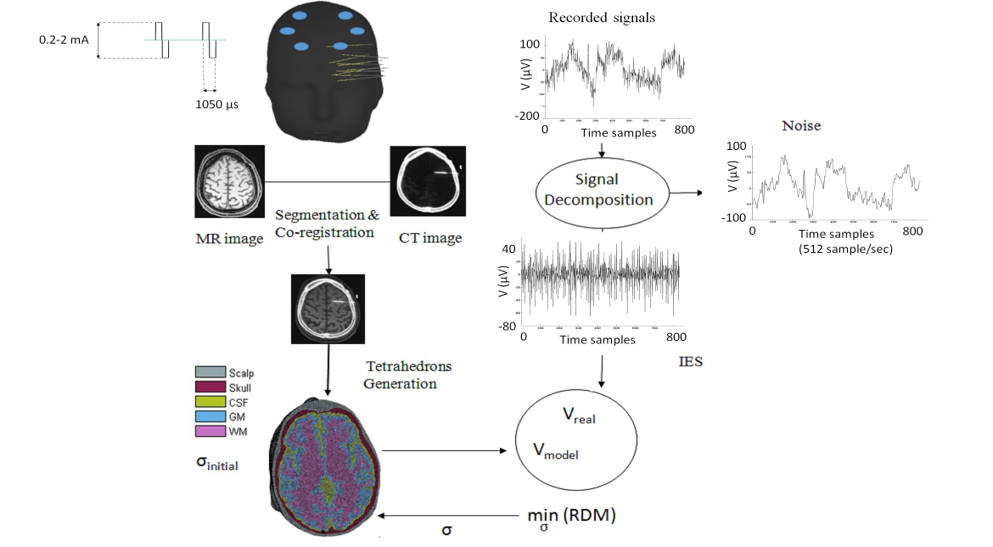}
\caption{The overall procedure of signal and image analyses for estimating head tissue conductivities}
\label{fig:overall}
\end{figure}

In-vivo conductivity estimation was performed by optimizing the conductivity values of the forward head model in order to fit the potentials acquired from the real patient as shown in Figure \ref{fig:overall}. For this purpose, homogeneous and isotropic five-compartments (scalp, skull, CSF, GM and WM) patient specific FEM head models were generated. For generating the head models, T1-weighted MRI and CT were first segmented. The segmentation of MRI was performed by the FreeSurfer software based on non-stationary anisotropic Markov Random Field (MRF), in which a probability of a label is modulated by the probability of its neighbors \cite{fischl2004sequence}. The segmentation of CT was based on intensity level thresholding. From the segmented CT, the positions of the intracerebral electrodes were determined by an automatic depth electrode localization algorithm \cite{hofmanis2011automatic}. After segmentation, as shown in Figure 1, MRI and CT were co-registered by maximizing the mutual information \cite{maes1997multimodality} to generate images segmented into five compartments: scalp, skull, CSF, GM and WM. In order to obtain a realistic head model solved by the numerical FEM method, the segmented images were then transformed into volume and this volume was then discretized by generating tetrahedrons \cite{fang2009Tetrahedral}. The tetrahedrons were generated by the TetGen program which is based on the Delaunay triangulation technique \cite{si2015tetgen}. By bounding the size of the tetrahedron to less than or equal to $1 mm^{3}$, the resulted means of the elements and the nodes over the three patients were: 1774481.33 and 306743.67 for scalp, 688385.67 and 132856.00 for skull, 724325.33 and 150504.33 for CSF, 672158.67 and 153916.33 for GM and 706215.67 and 135951 for WM. In homogeneous and isotropic head models, each $i^{th}$ compartment is assigned with a conductivity represented by a scalar value $\sigma_{i}$. In our setup, we have only considered 5 distinct conductivity values for the five compartments. These five values can be grouped in a conductivity vector $\bm{\sigma}$.   

The source is represented by the IES stimulation generated from two adjacent contacts of one intracerebral electrode. This (two monopoles) stimulation source can be modeled as a current dipole $j_{s}(t)$ situated at the midpoint between the two adjacent contacts, oriented along them and having a moment $I_{s}(t) \cdot d$, with $I_{s}(t)$ the injected current (biphasic pulses) and $d$ the distance between the stimulation contacts. For every patient and every stimulation position, the IES generated potential $x_{s}(\bm{\sigma})$ on every sEEG/EEG contact can thus be computed by solving the forward problem, using the specific head geometry and different conductivity values grouped in the vector $\bm{\sigma}$. Note that the measuring electrodes were also approximated as points, as justified in \cite{von2012electrode, caune2014evaluating, le2017seeg}.

\subsection{Signal Processing}

The goal of this part is to decompose sEEG-EEG measurements during electrical stimulation and to separate and denoise them to recover only the propagated electrical stimulation. From the acquired sEEG and EEG signals, only the time windows which correspond to the two-second stimulation period without commutation artifacts were considered. In order to have a valid approximation of the dipole source model and to avoid saturated recordings; the contacts which share the same electrode with the stimulating contacts were neglected.

Formally, the measured sEEG-EEG signals at a given time instant $\bm{x} = [x_{1}, \cdots ,x_{M}]^{T}$ from the M electrodes may be written as an instantaneous linear system:

\begin{equation}
\bm{x} = \bm{A}\bm{j_{p}} + \bm{n}
\end{equation}

where $\bm{A}$ $(M \times K)$ is the mixing lead-field matrix, $\bm{j}_{p}$ vector of $K$ physiological sources and $\bm{n}$ the noise (we have dropped the time index for simplicity, but the potentials vary in time because of the time variability of the sources). The propagated electrical stimulation signal $\bm{x}_{s}$ can be added to the previous model, such as:

\begin{equation}
\bm{x} = \bm{x_{s}} + \bm{x_{n}} + \bm{n} 
\end{equation}

where $\bm{x_{s}}$ is a vector containing the potentials (at time t) due to the propagated IES $\bm{j}_{s}(t)$ (high frequency, compared to the background physiological signals). The aim of the signal processing procedure described here is to extract $\bm{x}_{s}$ from the measured $\bm{x}$. As shown in \cite{hofmanis2013denoising} the best results for extracting the 55 Hz IES from the physiological background are obtained by a subspace decomposition procedure, briefly recalled in the next paragraph.

The proposed procedure is based on a combination of low pass filtering and Generalized Eigenvector Decomposition (GEVD). The main assumption is that, knowing the frequency characteristics of electrical stimulation, we can design a low-pass filter to apply on the raw sEEG signals $\bm{x}$ in order to obtain a second data set $\bm{y}$ containing mainly physiological sources while eliminating high frequency artifacts noise. Joint diagonalization of the covariance matrices of the raw and low-passed signals by GEVD (noted $\bm{R}_{x}$ and $\bm{R}_{y}$ respectively) simultaneously generates the eigenvectors for the raw versions of the sEEG/EEG measurements and their filtered version. Thus, after decomposition, the components recovered by high energy GEVD (i.e. high eigenvalues) will represent the common and therefore low frequency components. Even if the IES is very energetic compared to physiological activities, it is highly reduced by filtering, therefore barely present in the covariance matrix $\bm{R}_{y}$ and thus rejected by the joint diagonalization. Formally, the GEVD yields the eigenvectors $\bm{V}$ that solve the joint diagonalization problem:

\begin{eqnarray*}
\bm{V}^{T}\bm{R}_{y}\bm{V} = \bm{D} \\
\bm{V}^{T}\bm{R}_{x}\bm{V} = \bm{I}
\end{eqnarray*}

where $\bm{D}$ contains the eigenvalues in descending order and $\bm{I}$ is the identity matrix.  One can obtain next the spatially filtered components $\bm{z} = \bm{V}\bm{x}$, naturally ordered starting from the common components of the two data sets (i.e. low frequency physiological sources and trend), while the IES artifact sources will appear among the last components.

Different low-pass filters techniques were tested in the cited publication, with singular spectrum analysis showing the best performances. Moreover, according to \cite{hofmanis2013denoising}, the Singular Spectrum Analysis – Generalized Eigenvalue Decomposition (SSA-GEVD) has given a superior performance when compared with other methods for separating the stimulation from the background noise (based on Blind Source Separation).

The estimated “denoised” IES potentials without physiological background can then be obtained by reprojecting back only the high frequency components (the last two components, noted $\bm{z}_{s}$, as indicated in \cite{hofmanis2013denoising}):

\begin{equation}
\bm{x}_{d}(t) = \bm{V}^{T}\bm{z}_{s}(t)
\end{equation}

The estimated signals $\bm{x}_{d}(t)$ were next post-processed by time averaging of the simulation peaks, in order to obtain a denoised vector of IES potentials $\bm{x}_{d}$, to be compared with the (conductivities dependent) FEM modeled potentials $x_{s}(\bm{\sigma})$ introduced in the previous subsection.

\subsection{Error Function}

The optimal conductivity values were estimated by minimizing the error between the averaged denoised IES potentials $\bm{x}_{d}$ and the simulated potentials $\bm{x}_{s}$ (by forward problem with different conductivities $\bm{\sigma}$ (see previous section):

\begin{equation}
\min_{\bm{\sigma}} error(\bm{x}_{s}(\bm{\sigma}), \bm{x}_{d})
\end{equation}

Many functions can be applied to calculate the error between the modeled potentials and the real denoised potentials for estimating in-vivo conductivities. However, for evaluating the topological error, the relative difference measurement (RDM) \cite{meijs1989numerical}, shown in Eq. \ref{eq:rdm}, was considered in this study.

\begin{equation}
\label{eq:rdm}
RDM = \sqrt{\sum_{k=1}^{M} \left( \frac{x_{d,k}}{\|(\bm{x}_{d})\|_{2}} - \frac{x_{s,k}}{\|(\bm{x}_{s})\|_{2}} \right)^{2}} 
\end{equation}

where $k = 1 \cdots M$ is the index of the sEEG contacts, $\bm{\sigma}$ is a vector of conductivities (for the 5 tissues) and $\|\cdot\|_{2}$ is the l2-norm.

As it can be readily verified, the RDM produces an output between zero and two and it is related to the correlation between the two potential vectors $\bm{x}_{d}$ and $\bm{x}_{s}$. Unlike simple quadratic error functions, it is not affected by the difference in the potential amplitudes because of the normalization. This allows us to neglect the influence of the modeled IES current amplitude for a given stimulation and consider thus only the propagation coefficients (the corresponding lead-field matrix column). Although in the general case this error measure could be affected by some bias of one of the potentials (for example, an unknown non-null reference for the real measurements), in our case this is not a problem, because (1) the real signals are reconstructed using the high frequency components after SSA-GEVD (the reference potential is in principle in the physiological band, much lower) and (2) the $\bm{x}_{d}$ vector is constructed by time averaging (thus cancelling possible remaining reference bias). Obviously, the simulated potentials $\bm{x}_{s}$ are not affected by a reference bias.

Note finally that the topological error measured by the RDM has been considered in previous studies as a criterion for localization errors \cite{fernandez2011estimation, vorwerk2014guideline, dannhauer2011modeling, vallaghe2008global} [22, 43–45]. Besides, the RDM has produced more reliable results in sensitivity analysis when it was compared with the relative error \cite{altakroury2017vivo}.

\subsection{Optimization}

\begin{figure}[thpb]
\centering
\includegraphics[scale=0.4]{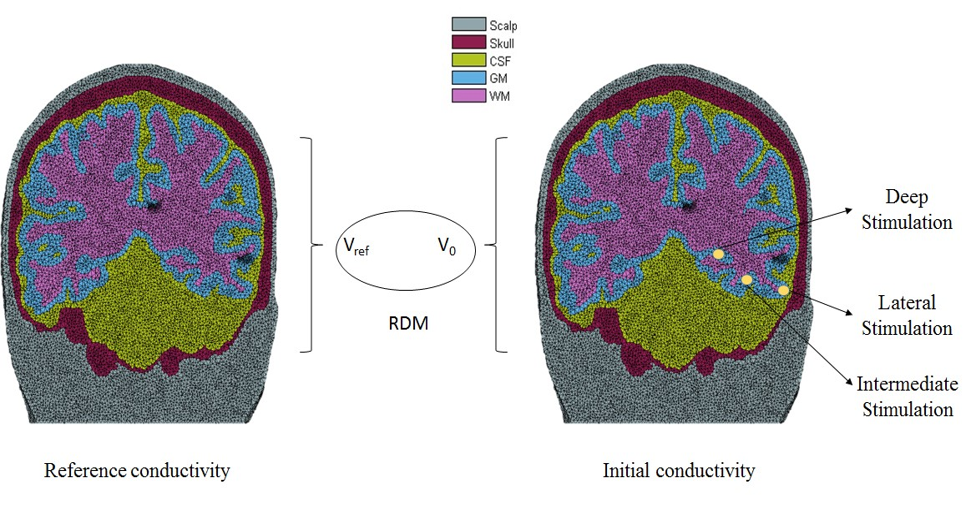}
\caption{The frontal view of two head models with the same geometry of Patient(1). The head model to the left acts as a reference head model, while the second to the right acts as a test head model}
\label{fig:test}
\end{figure}

In order to ensure that the optimization algorithm is robust, we performed a comparison by simulation among three different optimization algorithms: Nelder-Mead simplex (NMS), the genetic algorithm (GA) and the simulating annealing (SA). These algorithms were chosen from the free-derivative class since the numerical FEM method cannot be solved analytically. Besides, these algorithms are robust, easy to implement and common in the field of optimization. In addition, most of the previous studies which performed in-vivo conductivity estimation, have considered or recommended these algorithms \cite{lew2009improved, ferree2000regional, fernandez2011estimation, zhang2006estimation, lai2005estimation}. In this simulation, two identical head models of Patient(1) were generated as shown in Figure \ref{fig:test}, where numerical and spatial properties of the first head model were the same as the second head model except for the conductivity values. In the first head model, which acted as a reference model, fixed conductivity values were assigned. These conductivity values were chosen from the common values which are found in the literature: Scalp 0.33 S/m, skull 0.008 S/m, CSF 1.79 S/m, GM 0.33 S/m and WM 0.14 S/m \cite{geddes1967specific, baumann1997electrical, haueisen2002influence}, while the second head model had different initial conductivity values and acted as a test model. At each iteration, the conductivity vector of the second model is modified, the RDM is calculated and the algorithm searches for the optimal global solution. The challenge is to determine the most efficient algorithm able to recover the conductivity vector assigned to the reference model. The estimation of the reference conductivities was performed by the three different optimization algorithms, where each optimization algorithm performed the minimization for 36 different times given the following different conditions:

\begin{itemize}
\item Initial conductivities: Three different initial points.
\item Stimulation position: Deep, Intermediate and lateral stimulation.
\item Electrodes positions: Only sEEG measurements (107 measurements) and sEEG with EEG electrodes (107 + 19 measurements).
\item Additional Noise: The optimization algorithms were
tested without additional noise first, and then with additional white Gaussian
noise which makes the SNR of the generated potentials 80 dB.
\end{itemize}

The above conditions were considered to ensure that the performance of the optimization algorithm is independent of the starting point, the stimulation position and the considered measurements.

\subsection{Sensitivity Analysis}

In order to estimate in-vivo conductivities by optimizing the forward model, the error in conductivity values should have a notable effect on the output potentials and on the error function (in our case the RDM). In other words, the output should be sensitive to the conductivity values. The sensitivity analysis of conductivities permits us to evaluate the effect of conductivities on the output topological error when the other parameters of the head model (source position, measurement positions and the number of compartments) are fixed. In this work, we studied the effect of assigning erroneous conductivity values by simulation. The simulation was performed using the same two head models as described above, but in the test head model all the assigned conductivity values were equal to the reference conductivities values except one conductivity value which was set by multiplying its reference value by a factor between 0.5 and 1.5. Based on the assumption that the error in the reference conductivity is uniformly distributed around the reference conductivity $\sigma_{ref}$ as shown in Eq. \ref{eq:distrib}, the steps between the ratios were uniform and fixed as 0.1 to have a sufficient number of points.

\begin{equation}
\label{eq:distrib}
\sigma_{err} \backsim U(0.5\sigma_{ref},1.5\sigma_{ref})
\end{equation}

The range of [0.5,1.5] was chosen so that it covers the conductivity values which were reported in the literature \cite{haueisen2002influence, van2004influence}. To find out if RDM has the same sensitivity to a specific conductivity for different setups, the simulations were performed considering the following conditions on the different parameters:

\begin{itemize}
\item Number of compartments: Three compartments (the scalp, the skull and the brain) and five compartments (the scalp, the skull, the CSF, the GM and the WM). Considering two different head models with different number of compartments and thus different dimensions of the parameter optimization space. We aim to asses the ability of the method to estimate 3 or 5 conductivities according to the different configurations of the procedure and to examine the difference between their performances.

\item Stimulation location: Deep, intermediate and lateral stimulation positions. To have unbiased results, different stimulation positions were considered for performing the sensitivity analysis.

\item Measurements: sEEG (107 measurements), EEG (19 measurements) and sEEG+EEG (126 measurements). Since sEEG was not considered in previous studies for conductivity estimation, the effect of sEEG on conductivity estimation was examined by sensitivity analysis and by real conductivity estimation and was compared to the performance of scalp EEG.
\end{itemize}

\subsection{Conductivity Estimation}

In-vivo conductivity estimation of the scalp, the skull, the CSF, the GM and the WM was performed by optimizing the five-compartment forward head model of three drug-resistant epileptic patients based on the sEEG and EEG signals recorded simultaneously with IES. In the University Hospital of Nancy, the EEG and sEEG signals of the patients were recorded for five days on average. During this period, each patient was stimulated by IES at different anatomical positions. In this research we have selected the recordings that were acquired during the first day of recording in order to ensure the optimal impedance (50 ohms) of the EEG scalp electrodes. We have chosen signals corresponding to different deep, intermediate and lateral stimulation positions. For each patient, in-vivo conductivity estimations were done by considering: sEEG, EEG and sEEG+EEG signals to examine the effect of these measurements on estimating the conductivities of the different compartments. The procedure for in-vivo conductivity estimation is shown in Figure \ref{fig:overall} with the NMS as an optimization algorithm. The number of sEEG contacts and EEG electrodes for the patients were: 107 and 19 for Patient(1), 106 and 20 for Patient(2), and 157 and 24 for Patient(3).

\subsection{Source Localization}

In order to ensure that the conductivity estimation is robust and accurate, source localization of the real source positions (IES positions) was performed given the estimated conductivities and the real measurements. Then it was compared to the source localization results given by the common conductivity values of the literature. This comparison was performed for Patient(1) given the estimated conductivities which reported RDM of 0.25 or less. Only sEEG sensors were considered for solving the inverse problem of source localization: combined EEG/sEEG inverse problems are not trivial, because simultaneous recordings are seldom available, and even in this case, the number of sensors is very small and the spatial sampling is irregular for surface EEG. On the other hand, when using only sEEG sensors, the conductivities have a small impact on the source localization outcome, as the sensors are implanted in the same medium as the sources and simple models such as infinite homogeneous or one sphere provide good localization precision for dominant dipoles, as in the case of IES (see \cite{caune2014evaluating}).

Note that source estimation/localization algorithms minimize the (penalized) error function between the measured potentials and the modeled ones, as for the conductivity estimation above. However, two differences exist:

\begin{itemize}
\item The parameters to be optimized are the source characteristics (positions, orientations and amplitudes).

\item The function to be minimized is not the RDM (shown in Eq. \ref{eq:rdm}), but rather the L2 norm error function between the potentials regularized by some method depending on the given conditions. For example, for dipolar fitting or more generally sparse solutions (i.e. looking for a unique source or a small number of sources explaining the measurements), one aims to have the best possible fit between the sparse model (eventually with a unique source) and the data. On the contrary, for electrical source imaging (distributed sources), the regularization term is designed to favor one of the infinite number of solutions perfectly fitting the data (when more sources than sensors are estimated, the linear system is underdetermined).
\end{itemize}

In our case, especially after SSA/GEVD separation of the IES, the unique dipole is in principle the most adapted. We have thus looked for IES source by choosing the parameters $\bm{\theta}$ (position, orientation and amplitude) of a unique dipole ensuring the best fit between simulated potentials $\bm{x}_{s}(\bm{\theta})$ and $\bm{x}_{d}$ (note the dependency of the simulated potentials $\bm{x}_{s}$ on the dipole parameters $\bm{\theta}$ instead of the conductivities vector $\bm{\sigma}$). This fit was performed both using the common conductivities and the estimated ones, after the optimization procedure.

Unlike in \cite{caune2014evaluating}, we did not optimized the dipole parameters in a continuous space. Instead, we constructed a subsampled leadfield L, for discrete dipole positions in the gray matter. The distance separating two source positions was fixed as 10 mm. The number of resulting source positions was N = 1106 in the GM. The (M × 3N) leadfield matrix L contains the generated potentials at the recording positions from each elementary dipole source (unit amplitude), knowing that at each position, three different elementary dipoles are considered for the three different orientations.

Despite the a priori knowledge that the IES has a unique or at least dominant source, we have also evaluated source localization results using sLORETA \cite{pascual2011assessing}, which has been shown to produced zero errors in ideal noiseless conditions. Even though the IES source, which can be considered as a bipolar, is focal, we applied the sLORETA method to examine the focality of the source and the effect of the conductivity parameters on source localization by a distributed source method. We evaluated in this case the localization precision both in terms of position (i.e. the distance between the maximum energy source and the actual position of the IES) and in terms of volume. The volume of the source is quantified by two values:

\begin{itemize}
\item The number of mesh points having amplitudes within 1\% from the maximum amplitude
\item The volume occupied by these points, proportional to the product of the standard deviations along the principal axes of the cloud of these points (mathematically speaking, we have computed the product of the eigenvalues of the spatial covariance matrix of the high amplitude sources, less than 1\% below the maximum).
\end{itemize}

\section{Results}

\subsection{Optimization}

The optimization algorithms were compared in terms of speed and robustness (convergence to the required value). Given all the different conditions (stimulation position, measurement positions, and different initial points), the NMS outperformed both the GA and the SA in accuracy and speed. One sample of the results is shown in Table \ref{table:optimization} given the lateral stimulation and the sEEG measurements without additional white Gaussian noise. Because the NMS outperformed both the GA and the SA in simulation, it was considered in this study to minimize the RDM equation for estimating real in-vivo conductivities.

\subsection{Sensitivity Analysis}

\begin{figure}[thpb]
\centering
\includegraphics[scale=0.4]{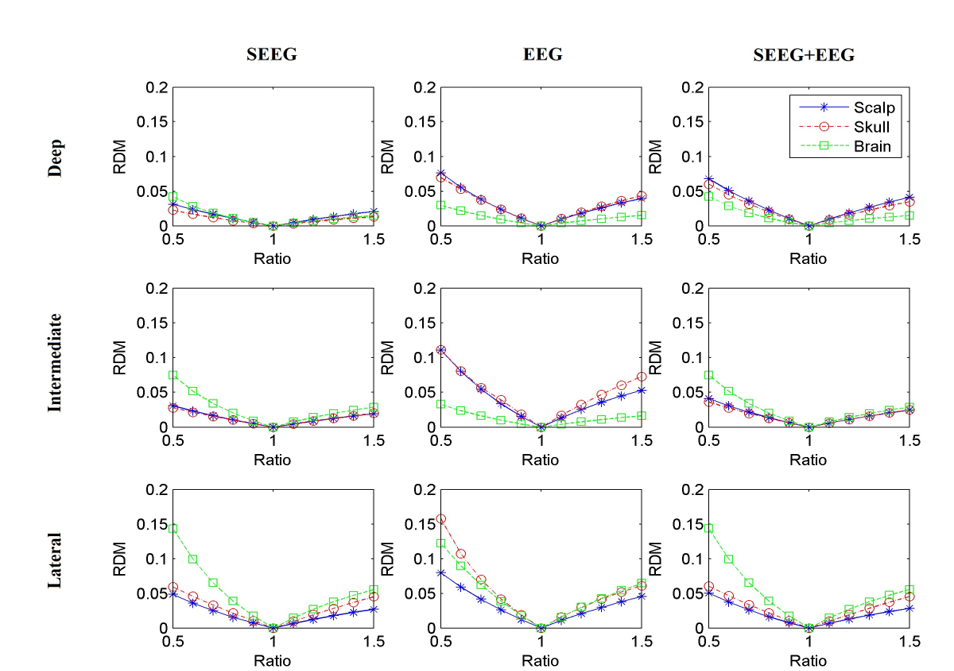}
\caption{The sensitivity of the RDM to the scalp, the skull and the brain conductivities in a three-compartment head model of Patient(1)}
\label{fig:sen3}
\end{figure}

\begin{figure}[thpb]
\centering
\includegraphics[scale=0.4]{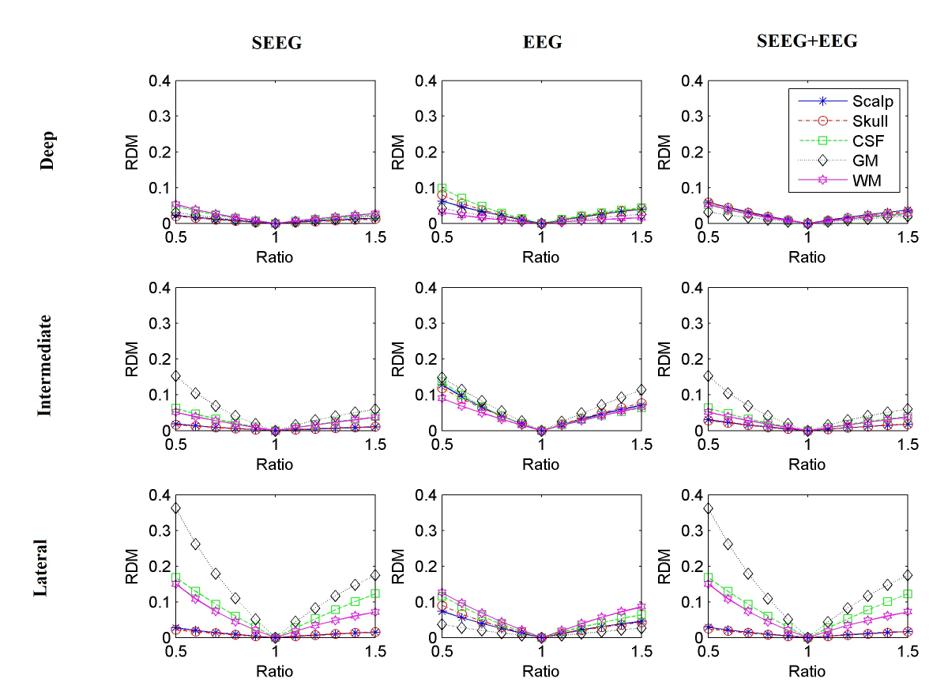}
\caption{The sensitivity of the RDM to the scalp, the skull, the CSF, the GM and the WM conductivities in a five-compartment head model of Patient(1)}
\label{fig:sen5}
\end{figure}

Sensitivity analysis was performed given deep, intermediate and lateral stimulation. For each stimulation position, the generated sEEG, EEG or sEEG+EEG measurements were considered. The results of sensitivity analysis in three-compartment head model and in five-compartment head model of Patient(1) are shown in Figure \ref{fig:sen3} and Figure \ref{fig:sen5} respectively.

\subsubsection{sEEG measurements simulation}

When considering only the sEEG measurements, it can be noted that the sensitivities to the scalp and the skull conductivities are the lowest in the three-compartment head model as shown in Figure \ref{fig:sen3} and in the five-compartment head model as shown in Figure \ref{fig:sen5} for all stimulation positions. The low sensitivities to the scalp and the skull conductivities are expected because the current generated by the IES does not propagate through these compartments before reaching the sEEG electrodes. Yet, in the three-compartment head model, as the stimulation is located laterally (closer to the scalp and the skull), the sensitivities to the scalp and the skull conductivities increase. But this increase is not significant in the five-compartment head model. On the contrary, in the five-compartment head model, the increase in the sensitivity to the deep compartments (especially the GM) is more remarkable than the increase in the sensitivity to the brain compartment in the three-compartment head model. This shows that there is a dependency between the conductivities assigned to the different compartments, because assigning one value to the brain compartment and varying the stimulation increases the sensitivities to the scalp and the skull, but when three different values were assigned for the brain compartment (as CSF, GM and WM) the effect of the stimulation position was not remarkable on the sensitivities of the scalp and the skull conductivities.

\subsubsection{EEG measurements simulation}

When considering only the EEG measurements, the sensitivities to the scalp and the skull conductivities in the three-compartment head model are higher than when considering the sEEG measurements as shown in Figure \ref{fig:sen3}. These higher sensitivities in the case of EEG are likely to occur because the potentials which are acquired by the EEG electrodes are due to field propagated through the scalp and the skull compartments before reaching the scalp electrodes, which is not the case for the sEEG electrodes situated below the skull compartment. Similarly, in the five-compartment head model, as shown in Figure \ref{fig:sen3}, the sensitivities to the scalp and the skull compartment in the case of EEG are higher than the case of sEEG. However, the difference between the sEEG measurements and the EEG measurement on the scalp and the skull conductivities is more notable in the three-compartment head model than the five-compartment head model. This difference can be explained by the dependency between the conductivities of the different compartments. Indeed, in the three-compartment head the scalp and the skull conductivities play a much more important role.

On the other hand, the sensitivity to brain conductivity in the three\-/compartment head model, as shown in Figure \ref{fig:sen3}, in the case of EEG measurement is lower than the sensitivity to the brain conductivity in the case of sEEG. The higher sensitivity to the brain compartment when considering the sEEG potentials is expected since the sEEG electrodes are nearer to the brain compartment than the EEG electrodes. However, when dividing the brain compartment into three different compartments (CSF, GM, and WM), as shown in Figure \ref{fig:sen5}, the difference between EEG and sEEG in the five-compartment head model is not as clear as it is in the three-compartment head model. This is due to the effect of the different stimulation on these deep compartments. When considering the EEG measurements, there is no remarkable difference between one stimulation and another on the sensitivity to the conductivities of the deep compartments. The sensitivities to these deep compartments are very similar to each other because the EEG measurements are not close to any of these compartments.

\subsubsection{sEEG+EEG measurements simulation}

When the EEG measurements are added to the sEEG, the sensitivity to the scalp and the skull conductivities in the three-compartment model as shown in Figure \ref{fig:sen3} becomes larger than when considering the sEEG measurements alone and lower than the case of EEG measurements alone for deep and intermediate stimulation. This sensitivity pattern is expected because the highest sensitivity to the scalp and the skull conductivities is likely to result when considering only the scalp EEG electrodes. Adding sEEG electrodes to EEG electrodes will reduce the sensitivity to the scalp and the skull conductivities, because the potentials do not cross the scalp and the skull potentials before reaching the sEEG electrodes. Yet, when considering the lateral stimulation, there is no remarkable change between the sEEG case and the sEEG+EEG case, because the larger distance between the lateral stimulation and the sEEG electrodes (compared to the intermediate and deep stimulation) reduces the effect of the sEEG electrodes in the sensitivity pattern. On the other hand, the sensitivity to the brain conductivity in the cases of sEEG+EEG and sEEG measurements are similar. This result is probably the most unexpected, indeed one would expect more differences between the two configurations of measurements. Adding EEG measurements to the sEEG did not alter the sensitivity to the brain conductivity because the number of the sEEG measurements is larger than the EEG measurements and because the EEG measurements are farther from the brain compartment than the sEEG measurements. Similarly, when considering the five-compartment model as shown in Figure \ref{fig:sen5}, the sensitivity to the conductivities of the deep compartments corresponding to the brain compartment (the CSF, the GM and the WM) does not change after adding the EEG measurements. However, the sensitivity to the scalp and the skull conductivities when considering the sEEG+EEG is larger than their sensitivity in the case of sEEG and lower than their sensitivity in the case of EEG in deep and intermediate stimulation as found in the three-compartment head model.

\subsection{Real Data Analysis}

\subsubsection{Conductivity estimation}

Real in-vivo conductivity estimation was performed for three drug-resistant epileptic patients given the sEEG and EEG signals recorded simultaneously with IES for each patient. Table \ref{table:pat1ex} shows the results of in-vivo conductivity estimation for Patient(1) considering the sEEG potentials. The table shows that the scalp conductivity value, in most of the cases, is equal to its upper bound fixed in the optimization which means that the estimate in this context cannot be considered. According to the sensitivity analysis presented above, these results were expected as all measurements are located before passing through the skull and scalp. When considering the RDM as the main criterion for stopping the optimization, it can be noted that the lower RDM (0.17) is obtained when the conductivity of the gray matter is equal to 0.13 S/m which is the nearest value among the other values to the common gray matter conductivity (0.33 S/m \cite{haueisen2002influence}). Having the greatest effect on the RDM from the gray matter conductivity is expected since the stimulation is situated in the gray matter. More interestingly, different estimated conductivity values (see Table \ref{table:pat1ex}) are obtained for different anatomical positions of the stimuli.

Since it is not possible to check the validation of the accuracy of the estimated conductivities in this study because we do not have real conductivity values for these three patients, the criterion we consider here is the variability among the estimated conductivities when they are estimated for the same position of the IES. The IES were classified according to their depth into three classes: deep, intermediate and lateral. The resulted means and variances of the conductivities are shown in Table \ref{table:pat1}, Table \ref{table:pat2} and Table \ref{table:pat3} for Patient(1), Patient(2) and Patient(3) respectively. It can be noted from the tables that most of the estimated scalp and skull conductivities are within the range of the estimated conductivity in the literature ([0.33-1] S/m for the scalp and [0.0042-0.05] S/m for the skull \cite{haueisen2002influence, van2004influence}), but the variability is high. Moreover, in accordance with the results of the sensitivity analysis previously presented, the variabilities of the estimated scalp and skull conductivities, in general, decrease as the stimulation goes more lateral or closer to these compartments. Regarding the skull conductivity, it can be noted from the tables that the resulted conductivities are lower when considering the EEG potentials than when considering only the sEEG potentials.

The importance of having the compartment between the stimulation and the measuring electrodes can be noted from the results of the CSF conductivities. In general, the estimated CSF conductivities are near the common value of the literature (around 1.79 S/m \cite{baumann1997electrical}) when considering the EEG measurements, since considering the EEG measurement allows the CSF compartment to be between the stimulation and the measurements. Similar results are not found in the EEG+sEEG case, since the number of the sEEG electrodes is much larger than the number of EEG electrodes and that biases the solution.

Most of the estimated GM and WM conductivities are within the literature common range for these-compartment ([0.33-1] S/m for the GM and [0.14-0.48] S/m for the WM \cite{haueisen2002influence, van2004influence}). The resulted variances of the estimated GM and WM conductivities agree with the sensitivity analysis results, since a high sensitivity indicates a low variance. This appears in the case of sEEG and EEG+sEEG for Patient(1) where the lowest variances are obtained when considering the lateral stimulation.

Because the amplitude of the stimulation decreases in proportion to the inverse of the distance square, having far measurements from the stimulation may reduce the accuracy of conductivity estimation and increase the variability of the estimated conductivities. In order to verify this hypothesis, an estimation was performed by considering only the measurements within 50 mm from the stimulation source. Table \ref{table:dist50} shows the results of conductivity estimation when considering the sEEG measurements within 50 mm from the stimulation source for Patient(1). It can be noted that the estimated scalp and skull conductivities are, in general, in the range of the values which are found in the literature ([0.33-1.0] S/m for the scalp and [0.0042-0.05] for the skull \cite{haueisen2002influence, van2004influence}). Although the estimated scalp conductivity in the case of deep stimulation has a zero variance, it cannot be considered since the estimated value is equal to the boundary of the optimization. However, the variances of the estimated scalp and the skull conductivities decrease as the stimulation goes more lateral as found in Table \ref{table:dist50}. Similarly, the estimated CSF conductivity in the case of deep stimulation is equal to the boundary of the optimization, so this value cannot be considered even though its variance is equal to zero. In general, the values of the estimated CSF are smaller than the common conductivity value (1.79 S/m \cite{baumann1997electrical}). This can be explained by the fact that the stimulation in addition to the measuring contacts are placed beneath the CSF compartment, so its conductivity value cannot be well-estimated. In spite of having a lower variance of the GM and the WM when considering the lateral stimulation (in agreement with the results found in Table \ref{table:pat1}), the estimated values of the lateral stimulation are out of the range of the conductivity values found in the literature ([0.33-1] S/m for the GM and [0.14-0.48] S/m for the WM \cite{haueisen2002influence, van2004influence}). However, the estimated GM and WM conductivities when considering the deep stimulation fit in the range of the values in the literature. This indicates the importance of having the stimulation near the compartment in order to estimate its conductivity. In general, the mean results found when considering the sEEG measurements within 50 mm distance are similar to the results found when considering all the sEEG measurements. However, there are changes in the values of standard deviations for some estimated conductivities (the skull, the GM and the WM) when considering measurements close to stimulation electrodes: they are clearly smaller.

\subsubsection{Source Localization}

One of the procedures impacted by the conductivity values is source localization. We have not carried on an extensive analysis of this problem in this paper, but we will present some preliminary results. More precisely, source localization of real IES positions was performed for Patient(1) given the estimated conductivities and the common conductivities which are found in the literature: Scalp 0.33 S/m, skull 0.008 S/m, CSF 1.79 S/m, GM 0.33 S/m and WM 0.14 S/m \cite{geddes1967specific, rush1968current, haueisen2002influence}.

Quantitative results of source localization are shown in Table \ref{table:local}, for three stimulation positions: deep, intermediate and more superficial. The results show that, when using distributed source localization procedures such as sLORETA, the performances are dependent on the conductivities and that the estimated conductivities lead to more precise source localization and spatial distribution (smaller source volume, here expected to be localized in a single point). On the other hand, the much better localization results obtained using estimated conductivities must be interpreted cautiously. Indeed, both conductivity estimation and source localization are mainly influenced by the nearby sensors and thus subject to overfitting. Nevertheless, regardless of the localization performance, the results show that the conductivity values are very important for distributed source localization approaches even when using only depth sensors (note that, for dominant dipole approaches, the conductivity values are far less important \cite{caune2014evaluating, le2017seeg}).

\section{Discussion}

Many studies have performed in-vivo conductivity estimation, but these studies did not succeed in providing the literature with conductivity values or ratios which can be used as references. The large variance among the estimated conductivity values is often attributed to the differences among the subjects in gender, age and health. The main purpose of this research was to show the effect of the source position and the measurement setup on conductivity estimation that can be achieved with: sEEG and EEG recorded in simultaneous with IES. These modalities, to the best of our knowledge, were not considered before for in-vivo conductivity estimation by optimizing the forward head model.

The forward head model acts as a basic building block for in-vivo conductivity estimation. If the head model is not accurate, the conductivity values must substitute the simplification in the head model and will not represent the real conductivity values. In this study, the effect of the model's parameters on the conductivity values was notable when comparing the sensitivity results of a five-compartment model with three-compartment model. Previous studies have found that having separate CSF, GM and WM compartments is important for building an accurate head model \cite{ramon2004role, vorwerk2014guideline}. In addition, it has been found that a segmented skull based on CT-scan gives a more accurate head model than a segmented skull based on the MRI \cite{montes2014influence}. Due to this, we considered a five-compartment head model consisting of scalp, skull, CSF, GM and WM from the MRI and the CT of each patient. Our model was chosen to be isotropic and homogeneous to reduce the computation time for conductivity estimation, and because the effect of inhomogeneity and anisotropy were found to be not as significant as having separate CSF, GM and WM compartments \cite{vorwerk2014guideline}. Still, the FEM head model can be extended in future studies to be inhomogeneous and anisotropic.

In-vivo conductivity estimation by optimizing the forward head model depends on having a robust optimization algorithm. Most previous studies which performed in-vivo conductivity estimation have considered the NMS algorithm for finding the minimum difference between the model potentials and the real potentials \cite{fernandez2011estimation, gonccalves2003vivo, zhang2006estimation, lai2005estimation}. However, other studies have applied or recommended other optimization algorithms like the GA and the SA \cite{lew2009improved, ferree2000regional}. In this study, we performed a comparison among the NMS, the GA and the SA for estimating conductivities. We found for different configurations in simulation that the NMS outperformed both the GA and the SA in speed and convergence to the solution. This urged us to consider the NMS in real in-vivo conductivity estimation. Even in real analysis, the NMS gave the same results for different initial values (multi-start approach). However, it should be noted that the NMS has less number of parameters compared to the GA and the SA. The additional parameters in the GA and the SA were set intuitively, but they were not validated by calculation. The small number of parameters in the NMS might have made it a preferable choice over the other optimization algorithms in the literature.

The results of the sensitivity analysis showed the importance of EEG scalp electrodes for in-vivo conductivity estimation of the scalp and the skull compartments. This was noted from the high sensitivity of the RDM to the scalp and the skull conductivities when the EEG potentials were considered. Similar results have been found by S. Vallaghé \cite{vallaghe2008global} and J. Haueisen et al. \cite{haueisen1997influence} who have concluded that the maximal effect on the EEG potentials comes from the skull and the scalp compartments which are located between the source and the measurement positions. The sensitivity of the RDM to the scalp and the skull conductivities was larger when considering the five-compartment FEM head model compared to the three-compartment FEM head model for the same measurement positions. This indicates that there is dependency between the conductivities of the different compartments. Besides, the higher sensitivity to the scalp and the skull conductivities, which resulted when the brain compartment was divided into three different compartments (CSF, GM and WM), is better for in-vivo conductivity estimation. In general, we found that the brain compartments affected the RDM when the sEEG measurement positions were considered, while the scalp and the skull affected the RDM when the scalp EEG was considered. Even though other studies have performed sensitivity analysis like the study of G. Marin \cite{marin1998influence} and the study of S. Vallaghé \cite{vallaghe2008global}, our study is the first which shows the effect of sEEG recordings on the sensitivity of the output potentials.

Due to clinical reasons, stimulations were not applied in one anatomical position, so the resulted conductivities had large variances. Our results would have low variances if all the stimulations had performed in one anatomical position but in this case the results would have been biased. Even though the results have large variances, in general, the mean values of the estimated conductivities were in the range of conductivities which are found in the literature \cite{haueisen2002influence, van2004influence}. Mainly, the effect of EEG scalp electrodes on the scalp and the skull conductivities was more remarkable than the effect of the sEEG electrodes on the deep compartments (CSF, GM and WM). This is because the EEG scalp electrodes are distributed uniformly on the scalp of the head while the intracerebral electrodes are implanted only in the region of interest. Ideally, intracerebral contacts should be distributed uniformly in the head in order to have the effect from the entire compartment.

In-vivo conductivity estimation was performed for three epileptic patients who had different positions of the epileptogenic zone and different number and positions of intracerebral electrodes (in addition to the difference in age and gender). Even though the estimated conductivities were not similar for these patients, general remarks were obtained. In general, the variabilities of the scalp and the skull conductivities were lower for lateral stimulations. The CSF conductivity had values closer to the values reported in the literature when the scalp EEG electrodes were considered. In addition, the skull conductivity was lower when considering only the EEG scalp electrodes. Getting a lower skull conductivity when considering the EEG electrodes is expected since the current has to pass through the high-resistive skull compartment before reaching the EEG electrodes. These results explain the low values of the brain-to-skull conductivity ratios (average 25 compared to the common ratio of 80 \cite{rush1968current}) which has been obtained in the research of Lai et al. who have considered both the scalp and the subdural potentials for estimating in-vivo conductivities \cite{lai2005estimation}. Furthermore, our results of source localization given the estimated conductivity values reported more accurate results than the literature conductivity values.

Previous studies have focused on the skull conductivity claiming that the CSF and the scalp conductivity values were well-estimated in the earlier ex-vivo studies \cite{lew2009improved, acar2016simultaneous}. However, their argument cannot be justified unless their head model represents the real head precisely. In-vivo conductivity estimation generates conductivity values which are adapted to the model simplification. Therefore, the conductivity values which are estimated in three-compartment model will not be equal to the values that are estimated in a five-compartment model as we showed in the sensitivity analysis. Due to this we did in-vivo conductivity estimation for all the compartments instead of considering one or two compartments.

\section{Conclusions}

With simultaneous intracerebral stimulations, depth (sEEG) and surface (EEG) recordings, we have shown that the variability among the estimated conductivities in the literature is not only due to inter-subject variability, but due to the differences in the source positions and measurement setups. The objective of this work is to highlight the confidence we can have in brain conductivity estimates and the limitations of the method in this context and these configurations. Indeed, the conditions and the spatial resolution of the measurements as well as the location of the source are very important parameters to take into account to ensure a better accuracy of the conductivity estimates. In other words, we have shown that the estimation of conductivities cannot be effective with a single stimulation location. Indeed, it is important to place the stimulation sites as close as possible to the tissues whose conductivities are to be estimated. Measurements must be distributed spatially near the sources but also on both sides of the brain areas. In other words, it is important that the electric field generated by the electric current source passes through the different structures and that the energy of the electric potentials collected by the electrodes on either side of the structures is sufficiently large to remain significant. Thus their weight becomes significant in the estimation of conductivities by multidimensional optimization under constraint.

From an application point of view, we have only evaluated the usefulness of conductivity estimation for source localization using sEEG electrodes. As shown in Caune et al. \cite{caune2014evaluating}, the location of dominant intracerebral sources from intracerebral measurements does not require brain models as complex as FEM models to locate the source(s) and it is not sensitive to accurate estimation of conductivity values. Nevertheless, if multiple (distributed) sources are assumed and thus the algorithmic approach to the inverse problem changes, the results are influenced by the conductivity values, even when using deep sensors only, both in terms of source location and volume. Although not treated here, one can thus suppose that EEG based distributed source localization (or mixed EEG/sEEG) will also be significantly influenced by the conductivity values. In this case the skull conductivity might play a very important role.

\begin{sidewaystable}[h]
\centering
\caption{The performance of optimization algorithms by simulation when considering a lateral stimulation with sEEG measurements without additional noise given the head model of Patient(1). The results for the different setups (measurement positions, stimulation positions and initial points) are found in \cite{altakroury2017vivo}}

\begin{tabular}{|c|c|c|c|c|c|} 
 \hline
 Algorithm & Initial Conductivity (S/m) & Resulted Conductivity (S/m) & Initial RDM & Resulted RDM & Time (Hours) \\ [0.5ex] 
\hline
NMS & First  & [0.33,0.0080,1.79,0.33,0.14] & $1.15*10^{-1}$ & $7.04*10^{-6}$ & 2.35 \\
    & Second & [0.33,0.0080,1.79,0.33,0.14] & $1.07*10^{-1}$ & $5.52*10^{-6}$ & 2.46 \\
    & Third  & [0.33,0.0080,1.79,0.33,0.14] & $6.36*10^{-1}$ &$1.40*10^{-5}$ & 2.47 \\ [0.5ex] \hline

GA  & First  & [0.33,0.0080,1.72,0.32,0.13] & $1.15*10^{-1}$ & $8.59*10^{-4}$ & 10.82 \\
    & Second & [0.33,0.0110,1.29,0.25,0.10] & $1.07*10^{-1}$ & $1.29*10^{-2}$ & 10.34 \\
    & Third  & [0.33,0.0171,0.63,0.18,0.06] & $6.36*10^{-1}$ &$6.02*10^{-2}$ & 10.33 \\ [0.5ex] \hline
    
SA  & First  & [0.33,0.0838,0.50,0.23,0.08] & $1.15*10^{-1}$ & $1.07*10^{-1}$ & 9.16 \\
    & Second & [0.33,0.0100,1.41,0.31,0.10] & $1.07*10^{-1}$ & $7.24*10^{-2}$ & 7.63 \\
    & Third  & [0.33,0.0730,12.48,2.08,0.86] & $6.36*10^{-1}$ &$7.56*10^{-2}$ & 9.23 \\ [0.5ex] \hline
 
\end{tabular}
\label{table:optimization}
\end{sidewaystable}
\begin{table}[h]
\centering
\caption{The estimated conductivities of Patient(1) given the sEEG measurements and a five-compartment head model.}
\begin{tabular}{|c|c c c c c|c|} 
 \hline
 IES position & \multicolumn{5}{|c|}{Resulted Conductivities (S/m)} & Resulted RDM  \\ 
 			  & Scalp & Skull & CSF & GM & WM 						& \\ [0.5ex] \hline 
 			  & 0.99 & 0.0082 &	0.18 & 0.64 & 0.10 & 0.23 \\
 			  & 0.22 & 0.0198 &	5.36 & 0.04 & 0.10 & 0.86 \\
Deep 		  & 0.99 & 0.0071 &	0.18 & 0.78 & 0.11 & 0.23 \\
 			  & 0.99 & 0.0240 &	0.18 & 0.05 & 0.07 & 0.75 \\
 			  & 0.99 & 0.0059 & 0.18 & 0.04 & 0.06 & 0.29 \\
 			  & 0.99 & 0.0097 &	0.18 & 0.06 & 0.06 & 0.18 \\ [0.5ex] \hline
 			  & 0.99 & 0.0050 &	0.18 & 0.13 & 0.37 & 0.17 \\
Intermediate  & 0.99 & 0.0100 &	0.18 & 0.04 & 0.08 & 0.22 \\
 			  & 0.03 & 0.0008 &	0.20 & 0.06 & 0.42 & 0.50 \\ [0.5ex] \hline
 			  & 0.99 & 0.0141 &	0.18 & 0.04 & 0.02 & 0.44 \\
Lateral 	  & 0.95 & 0.0240 & 0.23 & 0.05 & 0.01 & 0.40 \\
 			  & 0.87 & 0.0208 &	0.27 & 0.05 & 0.01 & 0.49 \\ [0.5ex] \hline
 			   
\end{tabular}
\label{table:pat1ex}
\end{table}

\begin{sidewaystable}[h]
\centering
\caption{The mean of the resulted conductivities (S/m) $\pm$ relative standard deviation (\%) of Patient(1)}
\begin{tabular}{|c|c|c|c c c c c|} 
\hline
Measurements & IES position & Number of stimulations & \multicolumn{5}{c|}{Resulted Conductivities} \\
             &              &                        & Scalp & Skull & CSF & GM & WM \\[0.5ex] \hline
             
     & Deep         & 6 & 0.54+46.4\% & 0.0047+200.0\% & 1.88+93.7\% & 0.47+76.6\% & 0.11+143.7\% \\
EEG  & Intermediate & 3 & 0.43+106.3\% & 0.0009+11.2\% & 2.70+37.0\% & 0.77+16.2\% & 0.25+77.0\% \\
     & Lateral      & 3 & 0.88+20.7\% & 0.0037+135.8\% & 3.77+73.6\% & 0.15+53.0\% & 0.04+107.1\% \\[0.5ex] \hline
     
     & Deep         & 6 & 0.86+36.6\% &	0.0125+60.6\% &	1.04+202.9\% & 0.27+128.8\% & 0.07+49.6\% \\
sEEG & Intermediate & 3 & 0.67+82.3\% &	0.0053+87.6\% &	  0.19+7.3\% & 0.08+68.3\% &	0.29+62.6\% \\
     & Lateral      & 3 & 0.94+6.6\%  & 0.0196+25.8\% &  0.23+20.5\% & 0.05+15.6\% &	0.02+13.9\% \\[0.5ex] \hline
     
     & Deep         & 6 & 0.60+56.8\% &	0.0152+58.9\% &	 0.22+32.8\% & 0.30+125.9\% & 0.14+95.6\% \\
EEG+sEEG & Intermediate & 3 & 0.31+77.9\% &	0.0084+78.5\% &	 0.24+30.5\% & 0.14+52.7\% &	0.40+6.3\% \\
     & Lateral      & 3 & 0.95+7.4\%  & 0.0209+26.0\% &  0.24+27.0\% & 0.05+18.0\% &	0.02+14.2\% \\[0.5ex] \hline
      
\end{tabular}
\label{table:pat1}
\end{sidewaystable}

\begin{sidewaystable}[h]
\centering
\caption{The mean of the resulted conductivities (S/m) $\pm$ relative standard deviation (\%) of Patient(2)}
\begin{tabular}{|c|c|c|c c c c c|} 
\hline
Measurements & IES position & Number of stimulations & \multicolumn{5}{c|}{Resulted Conductivities} \\
             &              &                        & Scalp & Skull & CSF & GM & WM \\[0.5ex] \hline
             
EEG      & Deep         & 13 & 0.63+46.3\% & 0.0046+135.8\% & 1.97+81.8\% & 0.37+95.7\% &	0.17+89.9\% \\
         & Intermediate & 4  & 0.73+33.3\% &  0.0093+37.3\% & 0.94+140.8\% & 0.39+105.8\% & 0.04+46.7\% \\[0.5ex] \hline
     
sEEG     & Deep         & 13 & 0.40+95.8\% &  0.0134+64.4\% & 1.01+163.6\% & 0.23+75.7\% & 0.25+63.1\% \\
         & Intermediate & 4  & 0.83+39.2\% &  0.0197+21.5\% &	0.18+0.0\%	& 0.50+92.9\% &	0.32+31.2\% \\[0.5ex] \hline
     
EEG+sEEG & Deep         & 13 & 0.53+75.7\% & 0.0096+71.3\% & 0.73+178.4\% & 0.23+73.8\% & 0.28+55.0\%\\
         & Intermediate & 4 & 0.87+27.6\% & 0.0187+21.7\% &	0.18+0.0\% &	0.47+91.3\% & 0.30+33.9\% \\[0.5ex] \hline
      
\end{tabular}
\label{table:pat2}
\end{sidewaystable}

\begin{sidewaystable}[h]
\centering
\caption{The mean of the resulted conductivities (S/m) $\pm$ relative standard deviation (\%) of Patient(3)}
\begin{tabular}{|c|c|c|c c c c c|} 
\hline
Measurements & IES position & Number of stimulations & \multicolumn{5}{c|}{Resulted Conductivities} \\
             &              &                        & Scalp & Skull & CSF & GM & WM \\[0.5ex] \hline
             
     & Deep         & 4 & 0.82+23.5\% & 0.0017+81.7\% &	2.79+75.8\% &	0.47+84.6\% & 0.27+72.0\%\\
EEG  & Intermediate & 4 & 0.87+13.9\% &	0.0022+67.9\% &	1.42+63.7\% &	0.54+70.1\% & 0.10+144.0\% \\
     & Lateral      & 4 & 0.99+0.0\% & 0.0008+7.4\% & 1.91+128.1\% & 0.51+108.0\% & 0.30+64.2\% \\[0.5ex] \hline
     
     & Deep         & 4 & 0.71+45.7\% &	0.0190+44.7\% &	0.32+51.0\% &	0.12+45.9\% & 0.08+56.2\% \\
sEEG & Intermediate & 4 & 0.20+66.4\% &	0.0190+40.5\% &	0.63+90.7\% &	0.18+63.4\% & 0.20+74.5\% \\
     & Lateral      & 4 & 0.59+80.4\% &	0.0155+63.4\% &	0.92+161.0\% & 0.05+72.5\% &	0.13+153.3\% \\[0.5ex] \hline
     
     & Deep         & 4 & 0.38+119.8\% & 0.0082+123.4\% & 1.12+86.3\% & 0.32+87.1\% &	0.24+74.9\% \\
EEG+sEEG & Intermediate & 4 & 0.21+72.7\% &	0.0183+42.3\% & 0.64+96.0\% & 0.19+68.5\% &	0.20+81.8\% \\
     & Lateral      & 4 & 0.81+19.4\% &	0.0127+83.2\% &	0.94+133.2\% & 0.29+158.3\% & 0.08+166.4\% \\[0.5ex] \hline
      
\end{tabular}
\label{table:pat3}
\end{sidewaystable}

\begin{sidewaystable}[h]
\centering
\caption{The mean of the resulted conductivities (S/m) $\pm$ the relative standard deviation (\%) of Patient(1) given his sEEG measurements within 50 mm of the stimulation}
\begin{tabular}{|c|c|c c c c c|} 
\hline
IES position & Number of stimulations & \multicolumn{5}{c|}{Resulted Conductivities} \\
              &                        & Scalp & Skull & CSF & GM & WM \\[0.5ex] \hline
             
Deep         & 6 & 0.99+00.0\% & 0.0093+83.6\% & 0.18+00.0\% &	0.57+73.4\% & 0.13+29.7\% \\
Intermediate & 3 & 0.88+22.1\% & 0.0082+23.9\% & 0.22+34.0\% &	0.10+31.7\% & 0.16+33.3\% \\
Lateral      & 3 & 0.93+11.2\% & 0.0197+23.1\% & 0.26+45.6\% &	0.06+23.3\% & 0.02+18.9\% \\[0.5ex] \hline

\end{tabular}
\label{table:dist50}
\end{sidewaystable}

\begin{sidewaystable}[h]
\centering
\caption{The impact of the estimated conductivity on the localization results where source localization quality is estimated using the localization error and the volume of the source. $\bm{\sigma_{r}}$ is the reference conductivity vector and $\bm{\widehat{\sigma}}$ is the estimated conductivity vector. For completeness, we also give the localization error using a unique dipole fit (the results are the same regardless of the conductivity \cite{caune2014evaluating})}
\begin{tabular}{|c|c c c c c|c|c|c|c||p{1cm}|} 
\hline
IES position & \multicolumn{5}{c|}{Resulted conductivities} & \multicolumn{4}{c||}{sLORETA} & DipFit loc. err. (mm) \\
      & Scalp & Skull & CSF & GM & WM & \multicolumn{2}{p{2.5cm}|}{Loc. err. (mm)} & \multicolumn{2}{p{4.5cm}||}{Scr. vol. (\# pts per vol.) $cm^{3}$}  & \\[1ex] 
 
 & & & & & & $\bm{\sigma_{r}}$ & $\bm{\widehat{\sigma}}$ & $\bm{\sigma_{r}}$ & $\bm{\widehat{\sigma}}$ & \\[0.5ex] \hline
 
Deep & 0.99 & 0.0097 & 0.18 & 0.06 &	0.06 &	6.49 &	14.74 &	7/0.65 & 641/106.11 & 6.49 \\
Intermediate & 0.99 & 0.0050 & 0.18 & 0.13 &	0.37 & 5.78 & 16.18 &	4/.040 & 72/8.72 & 5.78\\
Lateral & 0.99 &	0.0100 & 0.18 &	0.04 & 0.08 & 3.65 & 12.82 &	1/0 & 136/16.66 & 3.65 \\ \hline
\end{tabular}
\label{table:local}
\end{sidewaystable}

\bibliography{ref}

\end{document}